\documentclass{elsart}
\usepackage{natbib}
\usepackage{graphicx}
\usepackage{amsmath, amssymb}

\begin{document}

\begin{frontmatter}


\title{Space-irrelevant scaling law for\\
 fish school sizes}

\author{Hiro-Sato Niwa}
\ead{Hiro.S.Niwa@fra.affrc.go.jp}

\address{Behavioral Ecology Section,
National Research Institute of Fisheries Engineering,
Hasaki, Ibaraki 314-0421, Japan}

\begin{abstract}
Universal scaling in the power-law size distribution of pelagic fish
 schools is established.
The power-law exponent of size distributions is extracted through the
 data collapse.
The distribution depends on the school size only through the ratio of
 the size to the expected size of the schools an arbitrary individual
 engages in.
This expected size is linear in the ratio of the spatial population
 density of fish to the breakup rate of school.
By means of extensive numerical simulations,
 it is verified that
 the law is completely independent of the dimension of the space
 in which the fish move.
Besides the scaling analysis on school size distributions, the integrity
 of schools over extended periods of time is discussed.
\end{abstract}

\begin{keyword}
power law \sep
finite-size scaling \sep
pelagic fish \sep
school-size distribution \sep
space dimension
\end{keyword}

\end{frontmatter}


\section{Introduction}
Pelagic fish commonly cruise as a school.
As they migrate about in a limited space, interaction between schools
can occur so that two of them encounter and aggregate, or a large
school splits itself into two smaller schools, or more.
The continuation of these interaction processes of fission and fusion
may eventually lead to a fat-tailed school-size distribution.
\citet{Bonabeau-Dagorn95} found power law in school-size distributions
of tropical tuna.
The power-law distributions of school sizes are quite general for
pelagic fishes \citep{Niwa98}, and quantitative analyses are now in
progress (Bonabeau {et al.}, 1998, 1999; Niwa, 2003).

A model of school aggregation proposed by \citet{Bonabeau-Dagorn95} is
based on a physical model of particle aggregation, i.e. open systems
showing power-law cluster-size distributions \citep{Takayasu89}.
Bonabeau and Dagorn's (1995) model is briefly sketched as follows.
Simulated fish schools move between sites on coarse-grained zones of
space, and aggregate when they meet.
School splitting is replaced by (re)injection process:
a certain fraction of each school is separated from the school;
the fish that have left their school are reinjected as individuals
(i.e. one-sized schools) into the system, while the total number of
individuals is kept fixed.
\citet{Bonabeau-Dagorn95} obtained a truncated power law with exponent
$-3/2$ in the mean-field case of the model, in which schools can move
from any site to any other site.

The mean-field assumption might not always be adequate, because there
are spatial constraints on movement.
Although the ocean is three-dimensional, fish may not fully use their
spatial environment.
Pelagic fish movement generally takes place in a horizontal
(two-dimensional) space.
They are limited in depth by physiological constraints, and do not
dive into the deeps.
Additionally, pelagic fish schools are concentrated in the vicinity of
the front which is the contact zone and collision line of two oceanic
currents \citep{Uda38}.
For instance, skipjack schools, {\it Euthynnus vagans} (Lesson),
aggregate in the interfacial region between the cold subarctic and
warm subtropic waters \citep{Uda36}.
They are constrained to move effectively in a one-dimensional space.
In the presence of a fish aggregating device (FAD), which is a
drifting log or an artificial device designed to attract fish, pelagic
fish do not make full use of the three-dimensional oceanic space and
are concentrated in the vicinity of a FAD (a point), so that the space
dimension effectively decreases to less than one.

Bonabeau {et al.} (1998, 1999) predicted spatial effects,
i.e. the dimensional reduction that
the power-law exponent of fish school-size distributions is modified
from the mean-field case of their model when the dimension of the space
is taken into account:
the absolute value of the exponent, in a version of their model on a
$d$-dimensional lattice (schools hop to neighboring sites only),
decreases when the space dimension $d$ decreases.

Recently, \citet{Niwa03} analyzed some existing data
for various species in terms of a school-size histogram of the population
of fish.
$N$ fish swimming together form an $N$-sized school.
The school-size distribution $W(N)$ is proportional to the observed
number of $N$-sized schools;
the size histogram of the population is then represented by
$P(N) = NW(N)$,
which is proportional to the fraction of fish in $N$-sized schools to
total population.
He reported that the distribution $P$ is an exponentially thin-tailed
distribution, and, the distribution $W$ follows a power-law decay with
exponent $-1$ and is truncated at a cut-off size.
A simple stochastic-differential-equation model was proposed to explain
the observed power-law behavior, and the predictions of the model were
found to be consistent with empirical data.
A remarkable feature is ``scaling'':
all the empirical distributions collapse onto a single curve if the data
are plotted in terms of scaled coordinates with the mean value of a
histogram $P(N)$,
for all various species and for all environmental factors including the
above mentioned space dimension.
Note that the power-law distribution $W$ does not have a well-defined mean;
contrary the rapidly decreasing distribution $P$ has a well-defined mean.

The empirically determined power-law exponents of school-size
distributions for specific data sets range from 0.7 to 1.8
(Niwa, 1998; Bonabeau {et al.}, 1999).
Fat-tailed school-size distributions are necessarily truncated because
the population is finite.
This truncation of the power-law regime might lead to a ``wrong''
estimation of the exponent.
I make use of the data collapse to extract the ``right'' exponent.
I here propose a finite-size scaling (FSS) form \citep{Binder-Heermann88}
for the size distribution $W(N)$
on the assumption that the distribution decays with the truncated
power-law form with biologically universal exponent.
The power-law exponent of distribution $W$ is determined by
experimentally fitting the FSS relation to achieve the best data
collapse.
It is investigated by simulating the school system of pelagic fish
whether or not the dimension of the space in which fish swim is relevant
to the power-law exponent.
FSS for fish school sizes is elucidated through the competition between
two processes in the interacting school system: aggregation and
splitting of schools.

In addition, it is numerically investigated how long a school stays
together, i.e. neither merges with any other schools nor breaks up.
The behavioral algorithms governing school formation and dynamics have
been extensively studied [e.g. \citet{Inada-Kawachi02} and references
cited therein].
Studies of the integrity of schools, however, are very few
[e.g. \citet{Lester-etal85} or \citet{Bayliff88} for experimental
studies;
\citet{Niwa96} for a modeling approach].
Numerical simulations of fish-school aggregation suggest conjectures
about real situations that could be tested by observations.

\section{The Data}

The enlarged sets of the data in \citet{Niwa03} for school-size
distributions of pelagic fishes are analyzed (summarized in
Table~\ref{tab1}).
The ways of estimating school sizes of pelagic fishes were catch per set
by a purse seine or acoustic surveys.
Catch-per-set data are expressed in school weight (in metric tons).
Acoustic-survey data are expressed in dimensional size of a school
(e.g. vertical thickness in meters), which can be reduced to the
biomass in a school:
the school biomass is proportional to the vertical cross-section, the
square of vertical thickness, or the square of diameter of a school
(Squire, 1978; Anderson, 1981; Misund, 1993; Niwa, 1995;
Misund and Coetzee, 2000).
%
\begin{table}[tb]
\caption[database]{
{\bf Species analyzed.}
}
\vspace*{0.3cm}
\label{tab1}
\begin{center}
\renewcommand{\arraystretch}{1.1}
{
\begin{tabular*}{33.5pc}{llcl}
\noalign{\hrule height0.8pt}
& Species & cut-off size$^{\mbox{\scriptsize a}}$ & Data sources \\
\noalign{\hrule height0.8pt}
$\vartriangle$ & Northern anchovy &
 $15.67\cdot\Delta N$ & \citet{Smith70}$^{\mbox{\scriptsize c}}$.\\
& {\it Engraulis mordax} $^{\mbox{\scriptsize b}}$ && Acoustic survey\\
$\bullet$ & Japanese sardine & $6.45\cdot(\Delta N/2)$ to & \cite{Hara90}.\\
& {\it Sardinops melanosticta} & $23.77\cdot(\Delta N/2)$ & 22 acoustic surveys\\
$\square$ & Tropical tuna$^{\mbox{\scriptsize d}}$ & $11.69\cdot\Delta N$ &
 \citet{Bonabeau-etal99}.\\
& (free swimming) && Data from fisheries\\
$\lozenge$ & Tropical tuna$^{\mbox{\scriptsize d}}$ & $4.80\cdot\Delta N$ &
 \citet{Bonabeau-etal99}.\\
& (caught in the vicinity of FADs) && Data from fisheries\\
$\circ$ & Herring {\it Clupea harengus} & $7.24\cdot\Delta N$ to & \citet{Reid-etal00}.\\
&& $10.88\cdot\Delta N$ & 4 acoustic surveys\\
\noalign{\hrule height0.8pt}
\multicolumn{4}{l}{$^{\mbox{\scriptsize a}}$The cut-off size of
 power-law distribution is calculated by Eq.(9). $\Delta N$ denotes}\\
\multicolumn{4}{l}{the class width of frequency data of school sizes,
 and will be omitted.}\\
\multicolumn{4}{l}{$^{\mbox{\scriptsize b}}$possibly including {\it
 Trachurus symmetricus}, {\it Sarda chiliensis}, {\it Scomber japonicus},}\\
\multicolumn{4}{l}{ and {\it Sardinops sagax}.}\\
\multicolumn{4}{l}{$^{\mbox{\scriptsize c}}$Data are cited in
 \citet{Anderson81}.}\\
\multicolumn{4}{l}{$^{\mbox{\scriptsize d}}$Three species ({\it Thunnus
 albacares}, {\it Katsuwonus pelamis}, and {\it Thunnus obesus})}\\
\multicolumn{4}{l}{ are mixed.}\\
\end{tabular*}
}
\end{center}
\end{table}

Let $R$ be the dimensional size (vertical thickness or diameter
acoustically measured) of a steady moving $N$-sized school.
The following relation between dimensional and social sizes holds in a
statistical sense:
\begin{equation}
 R = (\mbox{constant})\times N^{\nu},
\end{equation}
where $\nu = 0.5$.
The prefactor is supposed to be constant for each data set (i.e. each
survey).
It may depend on the species and vary with regions, seasons and
years in which fish schools are surveyed.
The acoustic-survey data are transformed into a social size
histogram as follows
\begin{equation}
 W(N) \mbox{d}N
  \propto
  W(R) R^{1/\nu -1} \mbox{d}R,
\end{equation}
where $W$ denotes the school-size distribution density.

The data are given by the set
$\left\{\left.\left(N_i,W_i\right)\right| i=1,2,\ldots,n \right\}$.
$W_i \Delta N$ reads the frequency of school sizes which lie
 within the $i$-th class
$\left[N_i-\Delta N/2, N_i+\Delta N/2\right)$,
where $\Delta N$ denotes the class width
and the $i$-th class mark is given by
$N_i = (i-0.5)\cdot \Delta N$.
From now on, to simplify the expression,
we will omit $\Delta N$ (or $\Delta N/2$ for Japanese sardine) in
mathematical formulae for processing empirical data, that is,
such unit of school size is introduced as
$\Delta N =1$ ($\Delta N =2$ for Japanese sardine).
A school of unit size contains a certain number of individuals,
for instance,
$\Delta N$ should be a number of fish correspond to 1000 kilograms for
tropical tuna.

\begin{figure}[tb]
 \centering
 \includegraphics[width=8cm]{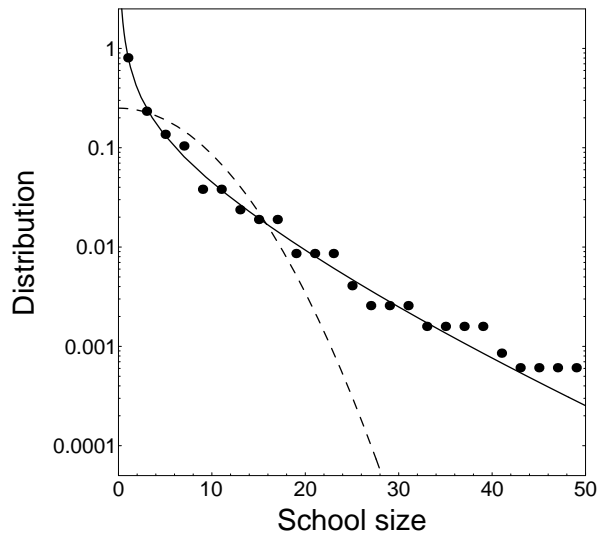}
 \caption{Comparison of the empirical data with Gaussian distribution
 (broken line).
 Japanese sardine schools were surveyed in the summer in 1982
 \citep{Hara90}.
 The trial Gaussian distribution has the variance $M_2$ equal to the
 second moment of empirical distribution
 ($M_2
 = \sum_i N_i^2 W_i / \sum_i W_i = 46.8$).
 The normalization of Eq.(17),
 \(
 \sum_{i=1}^n N_i W_i
 =
 \langle N \rangle_P
 \),
 is chosen,
 where $\langle N \rangle_P$ is given by Eq.(9).
 The Gaussian distribution is consistently normalized.
 The probability to find sardine schools ranging from 18 to 20 meters in
 vertical thickness (denoted by $p_{19}$) is $2.0\times 10^{-3}$ from
 the data.
 The Gaussian distribution gives $p_{19} = 8.1\times 10^{-10}$.
 The social size of such sardine schools as
 $R\in [18,20)$ in meters
 are transformed into
 $N\in [42, 50)$.
 The solid line shows the size distribution proposed by \citet{Niwa03},
 with $\langle N \rangle_P = 11.7$
 calculated from the data.
 Plotted on semilogarithmic scale.
}
 \label{fig:1}
\end{figure}

As with fisheries such as fishing by a purse seiner, such major events
(i.e. large catches) can be regarded as unusual events of great
magnitude lying on the tail of a distribution comprising events that are
mostly of much smaller magnitude.
Figure~\ref{fig:1} shows the school-size distribution for Japanese
sardine {\it Sardinops melanosticta},
which was from the acoustic survey off southeastern Hokkaido covering
the period July 30 -- August 6 in 1982 [\citet{Hara90}; the same data are
available in \citet{Hara84}].
A traditional, widely used Gauss statistics says that finding sardine
schools ranging from 18 to 20 meters in vertical thickness should only
occur about once every $10^9$ detections of schools.
In other words, it is not the real world!
Aquatic observations actually say that finding such schools occurs about
once every 500 detections.
The probability that such schools are found is $10^6$ times large!
%

\subsection{Finite-size Scaling}

The size distributions of pelagic fish schools follow a power law
$W(N) \propto N^{-\beta}$
up to a cut-off size
\begin{equation}
 L
  =
  \frac{
  \sum_{i=1}^n N_i^{1+\beta} W_i
  }{
  \sum_{i=1}^n N_i^{\beta} W_i
  }.
\end{equation}
In order to characterize the effects of the finite population size on
the truncation of power-law distribution,
a finite-size scaling hypothesis is used:
the distribution depends on $N$ only through the ratio $N/L^A$,
\begin{equation}
 \tilde{W}(N; L) \mbox{d}N
  =
  L^{-B}
  F \left(N/L^A\right) \mbox{d}\left(N/L^A\right),
  \label{FSS}
\end{equation}
where $F$ is a universal function independent of system (population)
size, and
$\tilde{W}_i = W_i/\sum_{i=1}^n N_i^{\beta} W_i$.
The prefactor $L^{-B}$ is required to ensure the normalization
\begin{equation}
 \sum_{i=1}^n N_i^{\beta} \tilde{W}_i = 1.
\end{equation}
School sizes are assumed to obey FSS with biologically universal exponents
$A$ and $B$ for a wide spectrum of both pelagic species and
environmental conditions.
The normalization Eq.(5) together with the postulated universal function
$F(x)$ gives
\begin{equation}
 \int_0^{\infty} L^{\beta A -B} x^{\beta} F(x)\mbox{d}x =1.
\end{equation}
It then follows that
\begin{equation}
 \beta = \frac BA,
\end{equation}
because all powers of $L$ must cancel out.
From the FSS hypothesis, it is expected that when
$
\tilde{W}(N) L^{A+B}
$
is plotted against
$N/L^A$
with correct parameters $A$ and $B$
all the empirical data should collapse onto a single curve.
The power-law exponent of fish school-size distributions, $\beta$, is
then evaluated through FSS analysis.
Besides Eq.(7), if FSS is valid the value of $A$ is $1$.
%

%
\begin{figure}[htbp]
 \centering
 \includegraphics[width=8cm]{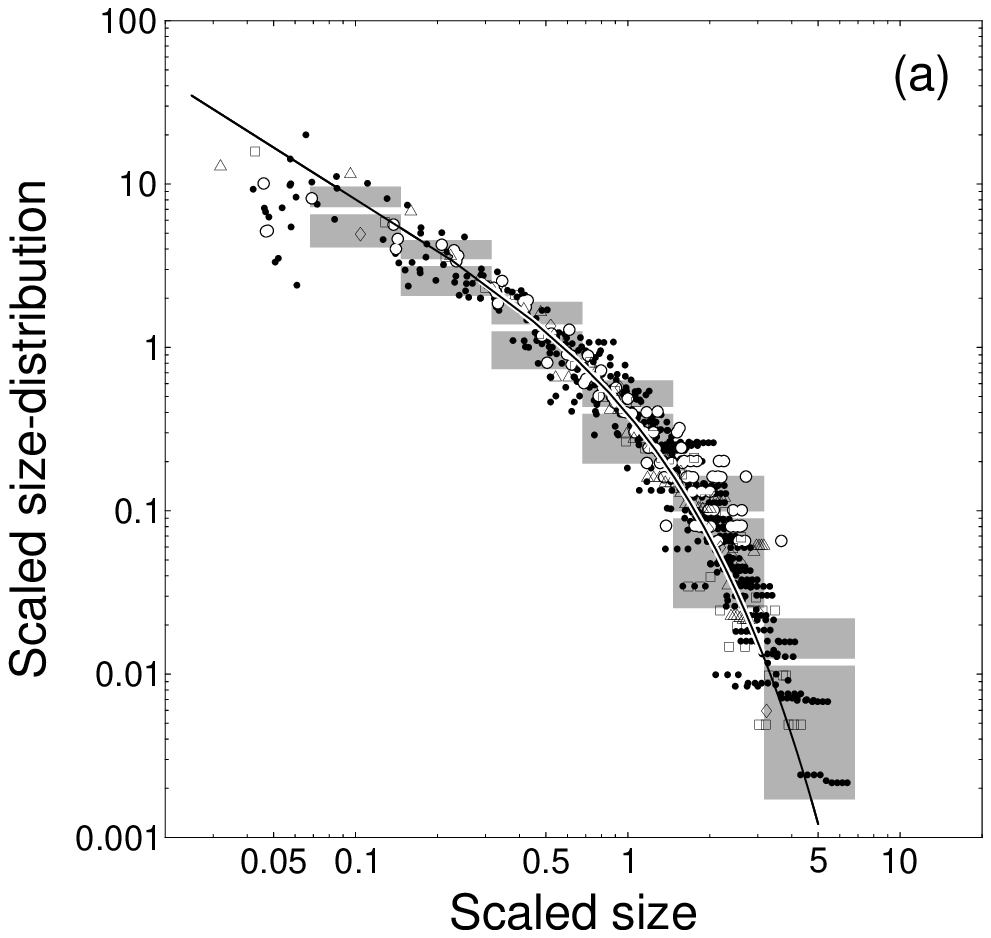}
 \includegraphics[width=8cm]{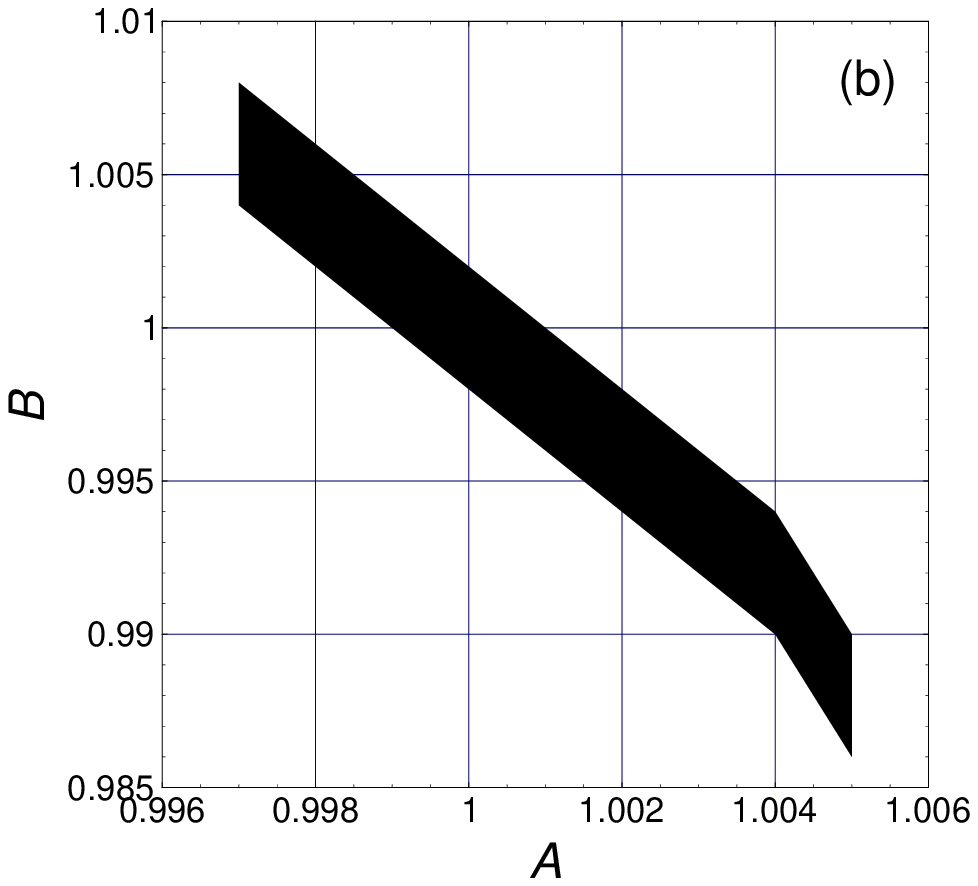}
 \caption{FSS analysis of the empirical data (summarized in
 Table~\ref{tab1}).
 (a)~FSS plot of the school-size distribution on double-logarithmic
 scale.
 Here $y= \tilde{W} L^{A + B}$
 is plotted versus
 $x= N/L^A$
 with $A =B =1$.
 The bins are chosen equally spaced on a logarithmic scale as
 $x \in \left[10^{-1+(k-0.5)/3}, 10^{-1+(k+0.5)/3}\right)$ with $k=0,1,\ldots,5$.
 For each bin two-dimensional variance $\epsilon$ is calculated
 (the rectangle in gray reads the interval $\overline{y}\pm\sigma_y$;
 the mean $\overline{y}$ is indicated by the slit).
 The solid line is a prediction of the mean-field theory \citep{Niwa03}.
 (b)~The region of the $AB$-plane in which
 the minimum of the mean of two-dimensional variance exists.
 The mean of two-dimensional variance, $\overline{\epsilon}$, takes a
 minimum $\overline{\epsilon}_{\mbox{\scriptsize min}}$
 for the right choice of $(A, B)$.
 The minimum is found with the precision i.e. width of the minimum,
 $\Delta\epsilon = 10^{-3}$ in black region
 ($
 \Delta\epsilon/\overline{\epsilon}_{\mbox{\scriptsize min}}
 \approx 2.58\times 10^{-3}
 $).
 The values of the parameters lie in the intervals
 $A = 1.001\pm 0.004$ and $B = 0.997\pm 0.011$.
 The value of $B/A$ is the estimate of the power-law exponent
 $\beta$, and therefore
 $\beta = 0.996\pm 0.015$.
 Experimentally fitting the parameters $A$ and $B$ to
 achieve a good data collapse yields the values
 $A = B = \beta = 1$, 
which are precisely the values of the mean-field case of the proposed
 model.
}
 \label{fig:2}
\end{figure}
Let us search for the values of $A$ and $B$ that do the best
job of placing all the data points on a single curve.
To do this, the $x$-axis is divided into bins (Fig.\ref{fig:2}a), and
the parameters are estimated at values that minimize the mean of
two-dimensional variance
\begin{equation}
 \epsilon
  =
  \left(\sigma_x/\overline{x}\right)^2
  +
  \left(\sigma_y/\overline{y}\right)^2
\end{equation}
(Lillo et al., 2002, 2003),
where $\sigma$ denotes the standard deviation, and $\overline{x}$ and
$\overline{y}$ denote the mean, and
the $x$-axis is chosen as
$x = N/L^A$,
and the $y$-axis represents
$
y = \tilde{W}(N) L^{A + B}
$.
The mean of two-dimensional variance,
$\overline{\epsilon}$,
is a measure to determine the goodness of collapse.
Figure~\ref{fig:2}b displays the set of pairs $(A, B)$
in which the minimum of the mean of two-dimensional variance,
$\overline{\epsilon}_{\mbox{\scriptsize min}}$, is guaranteed to lie.
A good data collapse can be obtained by using the values
$A \approx 1$ and $B \approx 1$.
The power-law exponent derived from the FSS collapse is $\beta \approx 1$. 
The resulting plot is shown in Fig.\ref{fig:2}a.
The figure confirms the FSS hypothesis, since all the data collapse
 onto a single curve.
The school-size distribution follows a power-law decay with exponent
$-1$, and is truncated at the cut-off size that equals the mean of
another distribution of the school system,
$P_i = N_i W_i$,
i.e. the size histogram of the fish population.
The distribution $P$ decays rapidly for large $N$, and has a well-defined
mean
\begin{equation}
 \langle N \rangle_P
  =
  \frac{
  \sum_{i=1}^n N_i^2 W_i
  }{
  \sum_{i=1}^n N_i W_i
  }.
\end{equation}

Two competing processes, mixing and splitting of schools, determine the
school-size distribution.
The spatial density of fish population conditions the mixing rate of
schools.
The population density of a given species differs among regions, seasons
and years.
The breakup rate of schools of a given species may vary depending on
environmental conditions,
e.g.
the lack of food may reduce school stability \citep{Morgan88}.
Figure~\ref{fig:2}a shows that the power-law exponent ``$-1$'' is robust,
while the cut-off size of linear power law varies (Table~\ref{tab1}).
The data-collapse implies that the breakup rate and the population
density do not affect the power-law exponent but the cut-off size.
This indicates that the power-law exponent is universal,
and hence the overall shape of the distribution may result from a simple
underlying aggregation mechanism.
Moreover, as a consequence of the quite good data-collapse, the
empirical data do not support the dimensional reduction, because the
``effective'' dimension is related to biological or environmental
conditions.
In the following two sections we will numerically examine whether or not
space dimension influences the power-law exponent.

\section{The Model}

A stationary equilibrium system of a fixed population size (number of
individuals, denoted by $\Phi$) is considered in which fish schools
break up and merge with other schools.
Let us investigate the aggregation process in a discretized space and
time.
There are $s$ sites on the lattice space $\Omega$.
On every site there is at most one school of simulated fish.
Let $N(j,t)$ be the size of the school on site $j$ at the $t$-th
time step.
At each discrete time step, each school hops to a new site or breaks
up into pairs of schools of various sizes possible.
If more than two schools happen to hop onto one site, they coalesce into
a single school with the size equal to the sum of the sizes of the
incident schools.
Assume binary splitting independent of school size.
Each school with a size greater than or equal to 2 splits into a pair of
schools with a probability $p$ at each time step.
The probability $p$ for a school to split per time step
(i.e. breakup rate) is independent of its size, and the sizes of
splitting schools are uniformly distributed:
a probability for an $N$-sized school to split into $M$- and
$(N-M)$-sized schools is represented by
\begin{equation}
 K_{\mbox{\scriptsize b}}(N|M,N-M)
 =
 K_{\mbox{\scriptsize b}}(N)
 =
  \frac{p}{N-1},
\end{equation}
for $N \geq 2$.

The aggregation process can be represented by the following stochastic
equation for $N(j,t)$:
\begin{equation}
 N(j,t +1)
  =
  \sum_{k\in\Omega} M_{jk} (t) N(k,t),
\end{equation}
where $M_{jk} (t)$ is a stochastic variable.
If the school on site $k$ does not break,
$M_{jk} (t)$ is equal to 1 when the school jumps to site $j$ and equal
to 0 otherwise
($M_{ik} (t) = 0$ for $i\neq j$).
If an $N$-sized school on site $k$ breaks up into two schools of sizes
$mN$ and $(1-m)N$ jumping to sites $j$ and $j'$, respectively,
then we have
$M_{jk} (t) = m$,
$M_{j'k} (t) = 1-m$, and the others vanish
($M_{ik} (t) = 0$ for $i\neq j,j'$),
where $m$ is also a stochastic variable uniformly distributed in
$(0,1)$ on condition that $mN$ is an integer.
$M_{jk} (t)$ must be normalized as
\begin{equation}
 \sum_{j\in\Omega} M_{jk} (t) = 1,
\end{equation}
for $\forall k$ and $\forall t$,
which guarantees the conservation of population.

In the mean-field case of the model,
at each time step, all schools move towards a randomly selected site,
which corresponds to the migration of the high potential speed of fish,
e.g. free-swimming tuna.
They may move to any site with equal probability $1/s$.
In spatial model of school aggregation, schools hop to neighboring sites
only.
In a version of the model on a $d$-dimensional lattice, they may move to
each of $2d$ neighboring sites with equal probability $1/2d$.
If a school on site $j$ breaks, one of splitting schools remains at the
site $j$ and the other hops to neighboring sites.

%
\begin{table}[tb]
\caption[simulation]{
 {\bf Parameters used in simulations.}
Monte-Carlo simulations of fish school aggregation on the lattice space
 of $s=2^{18}$ sites with periodic boundary have been conducted by using
Mersenne Twister \citep{Matsumoto-Nishimura98}, a pseudorandom number
 generator, on Scientific Computing System of MAFFIN, Tsukuba Japan.
}
\vspace*{0.3cm}
\label{tab2}
\begin{center}
\renewcommand{\arraystretch}{1.1}
{
\begin{tabular*}{27pc}{lccc}
\noalign{\hrule height0.8pt}
& breakup rate $p$ & population $\Phi$ &
 $\langle N \rangle_P$\ \ [1D;\ 2D;\ mf]\ $^{\mbox{\scriptsize a}}$\\
\noalign{\hrule height0.8pt}
$\blacklozenge$ & 0.02 & $2^{14}$ & 1.72;\ 4.23;\ 8.97\\
$\blacksquare$ & 0.02 & $2^{15}$ & 3.08;\ 8.57;\ 17.87\\
$\blacktriangle$ &0.02 & $2^{16}$ & 6.39;\ 18.57; 38.40\\
$\bullet$ & 0.02 & $2^{17}$ & 13.44;\ 39.41;\ 74.25\\
$\bigstar$ & 0.02 & $2^{18}$ & 28.39;\ 73.83;\ 157.61\\
\hline
$\lozenge$ & 0.01 & $2^{17}$ & 19.63;\ 68.59;\ 146.61\\
$\square$ & 0.03 & $2^{17}$ & 10.98;\ 27.00;\ 48.51\\
$\vartriangle$ & 0.04 & $2^{17}$ & 9.13;\ 20.29;\ 37.03\\
$\circ$ & 0.1 & $2^{17}$ & 5.54;\ 9.41;\ 13.94\\
\noalign{\hrule height0.8pt}
\multicolumn{4}{l}{$^{\mbox{\scriptsize a}}$
 $\langle N \rangle_P$ was computed from simulation results after
 $2^{17}$ time steps}\\
 \multicolumn{4}{l}{for one- (1D), two-dimensional (2D), and
 mean-field (mf) cases.}\\
\end{tabular*}
}
\end{center}
\end{table}

\subsection{Finite-size Scaling}

Simulations have been performed with the coarse-grained zones of
$s=2^{18}$ sites, simulation run $=2^{17}$ time steps, and parameters
summarized in Table~\ref{tab2}.
The initial school-system configurations are taken to be random
distribution of eight-sized schools on the lattice space.
Numerical results are shown in the next section
(in Fig.\ref{fig:6}b an FSS collapse of $W(N)$ for the two-dimensional
case is depicted).
We search for the values of $A$ and $B$ that place all of the simulated
distributions most accurately on a single curve.
The parameters $A$ and $B$ derived via the minimization of the
measure $\overline{\epsilon}$ to quantify FSS collapse are summarized in
Table~\ref{tab3}.
In the one- and two-dimensional cases, there is a clear minimum for
$A \approx B \approx 1$.
Therefore, the power-law exponent extracted from numerical simulations
reads $\beta \approx 1$, and does not depend on the dimensionality $d$.
This may cause one surprise,
because such exponents depend on the dimensionality as
the critical exponents for scaling behavior in wide varieties of
physical phenomena, e.g. magnetization, specific heat, size of a
polymer, and so on.
%
\begin{table}[tb]
\caption[simulation]{
 {\bf Summary of exponents $A$, $B$, and $\beta$.}
 The exponents are evaluated at values in the listed ranges
 with the precision $\Delta\epsilon = 10^{-3}$
 through FSS analyses of empirical data and simulated distributions at
 the last of run for different population sizes and breakup rates in
 one- and two-dimensional cases.
}
\vspace*{0.3cm}
\label{tab3}
\begin{center}
\renewcommand{\arraystretch}{1.2}
{
\begin{tabular*}{29pc}{lcccc}
\noalign{\hrule height0.8pt}
 & $A$ & $B$ & $\beta$ &
 $\Delta\epsilon/\overline{\epsilon}_{\mbox{\scriptsize min}}$\\
\noalign{\hrule height0.8pt}
data & $1.001\pm 0.004$ & $0.997\pm 0.011$ & $0.996\pm 0.015$ &
 $2.58\times 10^{-3}$\\
1D & $0.998\pm 0.012$ & $1.020\pm 0.026$ & $1.022\pm 0.037$ &
 $2.93\times 10^{-3}$\\
2D & $0.992\pm 0.011$ & $1.019\pm 0.026$ & $1.028\pm 0.037$ &
 $1.01\times 10^{-2}$\\
mf\/$^{\mbox{\scriptsize a}}$ & 1 & 1 & 1 & ---\\
\noalign{\hrule height0.8pt}
\multicolumn{4}{l}{$^{\mbox{\scriptsize a}}$
 predicted values of the mean-field theory \citep{Niwa03}. 
}\\
\end{tabular*}
}
\end{center}
\end{table}

\section{The Scaling Law}

\begin{figure}[tb]
 \centering
 \includegraphics[width=8cm]{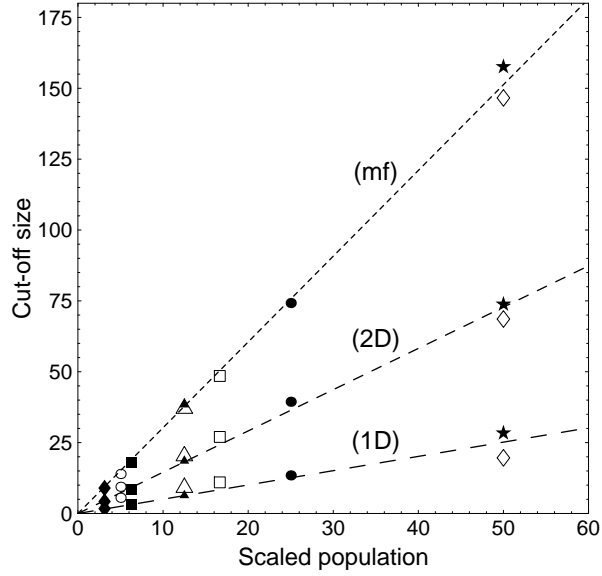}
 \caption{Plot of $\langle N \rangle_P$ against the scaled population
 size (numerical results from simulations summarized in
 Table~\ref{tab2}).
 The abscissa represents $\rho/p$,
 the coarse-grained population density $\rho\;(=\Phi/s)$ divided by
 breakup rate $p$.
 Broken lines are fits with slopes
 0.503 (1D),
 1.46 (2D), and
 3.03 (mf),
 respectively,
 which give the prefactor $c$ in Eq.(13).
}
 \label{fig:3}
\end{figure}

The cut-off size of power-law distribution
(equal to $\langle N \rangle_P$)
results from variable individual behavior (i.e. breakup rate $p$) and
fluctuating population density (denoted by $\rho$).
The FSS collapse suggests that the school-size distribution $W(N)$
depends on $N$, $p$, and $\rho$ only through the variable
$x = N/\langle N \rangle_P$.
Numerical simulations reveal that $\langle N \rangle_P$ depends linearly
on the spatial population density and inversely on the breakup rate
(Fig.\ref{fig:3}):
\begin{equation}
 \langle N\rangle_P = c\rho/p.
\end{equation}
The prefactor $c$ must have dimensions of
$[\mbox{length}]^d/[\mbox{time}]$,
and it is indeed proportional to the ratio of the coalesce rate to the
spatial density of schools, as discussed later.
%

\begin{figure}[tb]
 \centering
 \includegraphics[width=8cm]{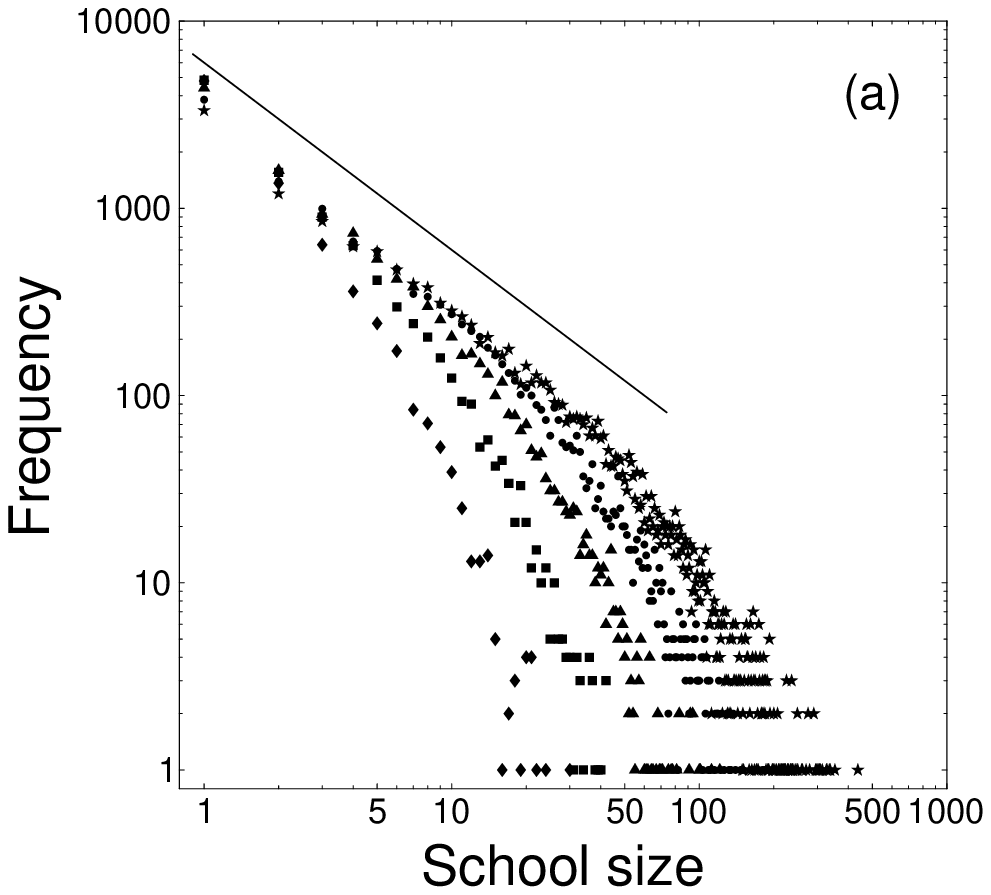}
 \includegraphics[width=8cm]{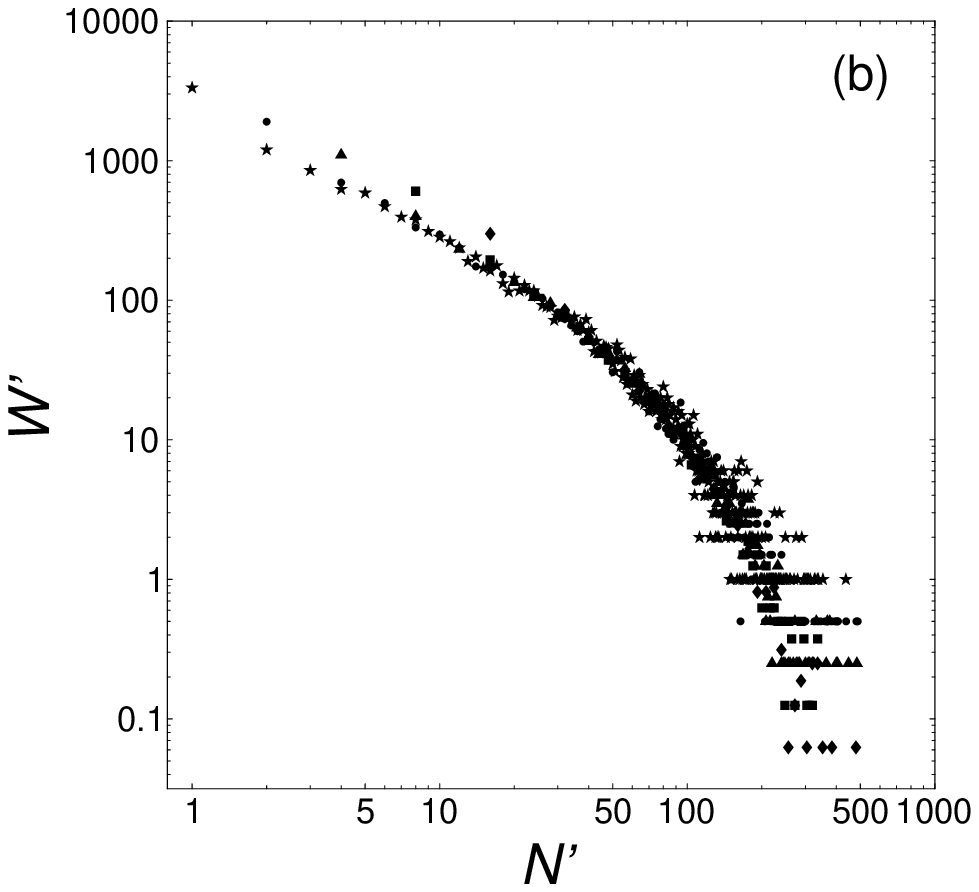}
 \caption{Simulated school-size distribution (2D) with the breakup rate
 $p=0.02$, $s=2^{18}$ sites after $2^{17}$ time steps.
 (a) Frequencies ${\hat W}(N)$ of $N$-sized schools in the
 two-dimensional simulations with various population sizes summarized in
 Table~\ref{tab2}.
 The straight line illustrates a general trend, i.e. the power-law
 behavior with exponent $-1$.
 (b) The frequencies in Fig.\ref{fig:4}a re-plotted with
 $W'={\hat W}(N)\rho$
 as a function of variable $N'=N/\rho$.
 The scaling causes a shift of the curves in Fig.\ref{fig:4}a that
 depend on $\rho$.
 All the distribution collapse onto a single curve.
 Plotted on double-logarithmic scale.
 }
 \label{fig:4}
\end{figure}

\begin{figure}[tb]
 \centering
 \includegraphics[width=8cm]{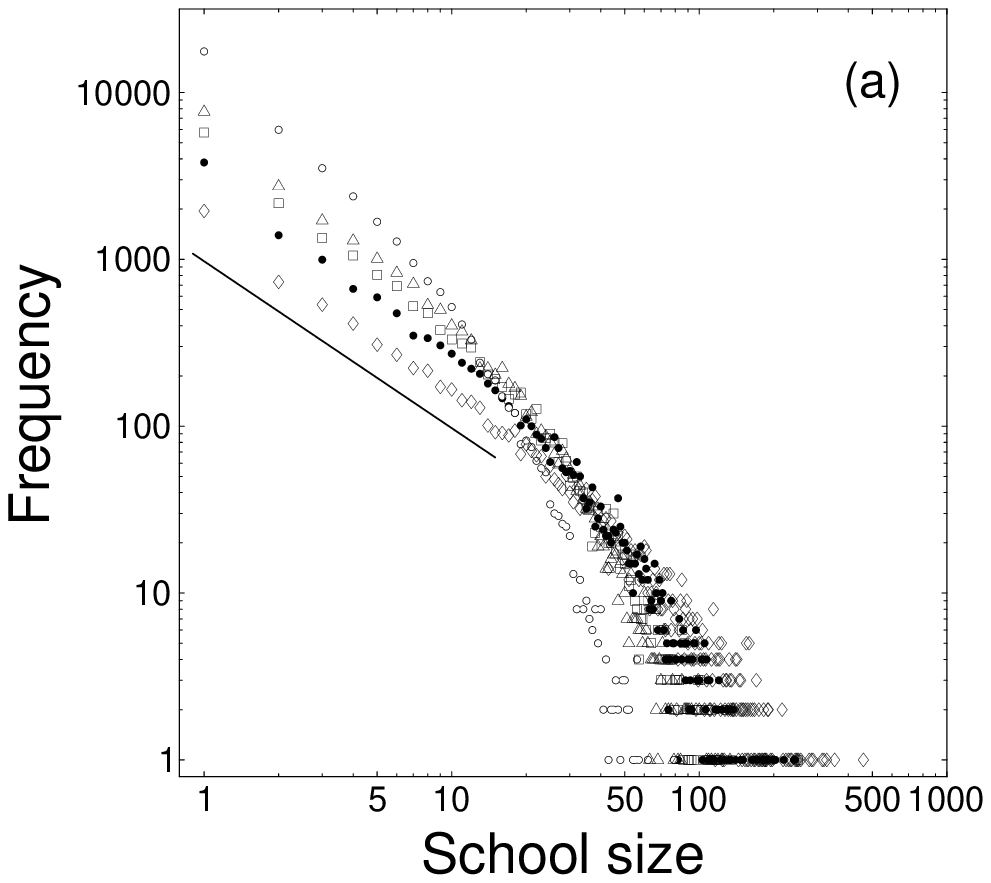}
 \includegraphics[width=8cm]{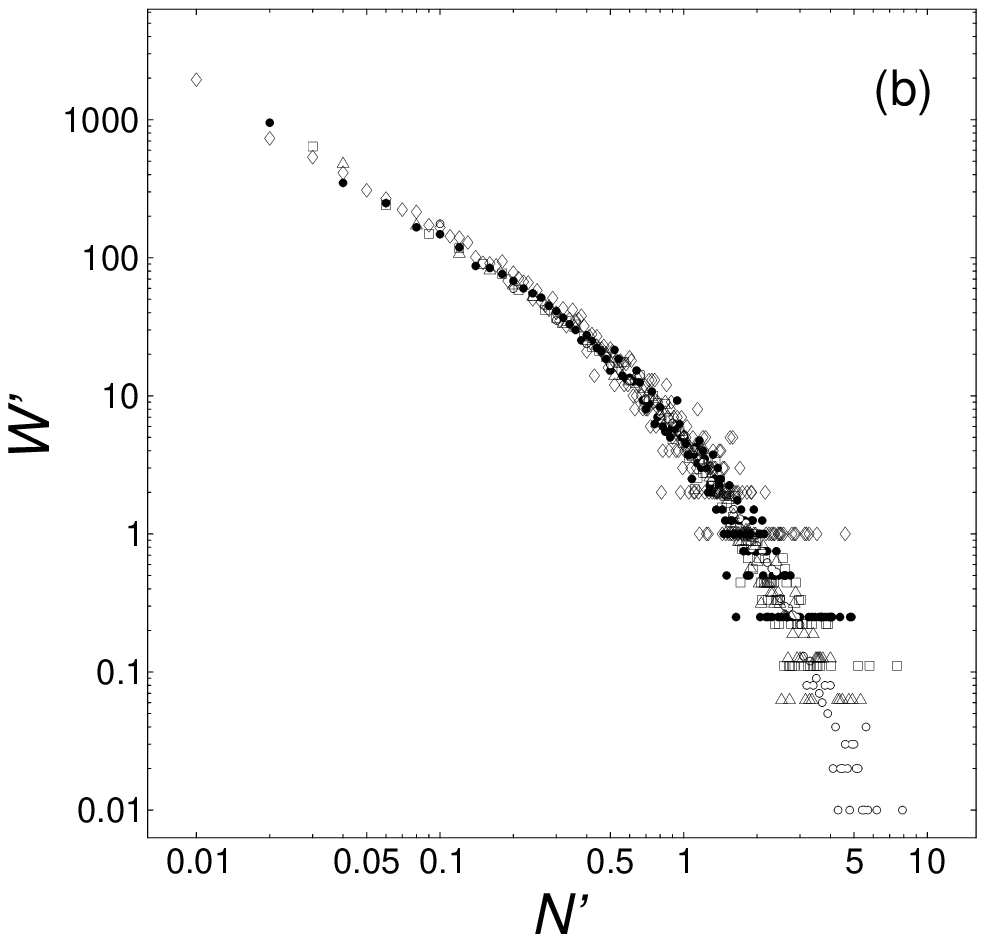}
 \caption{Simulated school-size distribution (2D) with the population
 density $\rho=0.5$ (population $\Phi =2^{17}$), $s=2^{18}$ sites
 after $2^{17}$ time steps.
 (a) Frequencies ${\hat W}(N)$ of $N$-sized schools in the
 two-dimensional simulations with various breakup rates summarized in
 Table~\ref{tab2}.
 The straight line illustrates a general trend, i.e. the power-law
 behavior with exponent $-1$.
 (b)~The frequencies in Fig.\ref{fig:5}a re-plotted with
 $W'=10^{-4}{\hat W}(N)p^{-2}$
 as a function of variable $N'=N p$.
 The scaling causes a shift of the curves in Fig.\ref{fig:5}a that
 depend on $p$.
 All the distribution collapse onto a single curve.
 Plotted on double-logarithmic scale.
}
 \label{fig:5}
\end{figure}

The frequencies of the amount of $N$-sized schools ($N=1,2,\ldots,\Phi$)
at the last of simulation run in the two-dimensional case,
${\hat W}(N;\rho,p)$, are shown in Figs.\ref{fig:4}a
and~\ref{fig:5}a for a given breakup rate $p$ and a given
coarse-grained spatial population density $\rho =\Phi/s$,
respectively.
The frequencies are normalized as
\begin{equation}
\sum_{N=1}^{\Phi} N{\hat W}(N) = \Phi.
\end{equation}
%
%

Examining Fig.\ref{fig:4}a closely shows that for fixed breakup rate
$p$ but increasing population density $\rho$,
the range of the power-law regime increases:
the cut-off size of power law scales with $\rho$ as depicted in
Fig.\ref{fig:4}b.
The FSS form~(\ref{FSS}) with $A =B =1$ together with Eq.(13)
reads
\begin{equation}
 {\hat W}(N;\rho,p=0.02) \mbox{d}N
  =
  N^{-1} g(N/\rho) \mbox{d}\left(N/\rho\right),
\end{equation}
where $g(x)$ is a scaling function.

Similar behavior is seen in Fig.\ref{fig:5}a when $p$ varies.
As $p$ increases, the length of the power-law region reduces:
the cut-off size of power law scales with $1/p$ as depicted in
Fig.\ref{fig:5}b, i.e.
\begin{equation}
 {\hat W}(N;\rho=0.5,p) \mbox{d}N
  =
  p N^{-1} g'(Np) \mbox{d}\left(Np\right),
\end{equation}
where $g'(x)$ is a scaling function.
Each resulting graph shows that all the distributions collapse onto a
single curve (Figs.\ref{fig:4}b and~\ref{fig:5}b).

The same results can be obtained with one-dimensional case of the
model, and its mean-field case as well.

\subsection{The Unified School-size Law}

Equation~(13) implies that
the population density must appear as a ratio to the breakup rate in
the scaling analysis, and the breakup rate vice versa.
For instance,
the following normalization is appropriate for the scaling
analysis of school based data:
\begin{equation}
 \sum_{i=1}^n N_i W_i
  =
  \langle N \rangle_P.
\end{equation}
The FSS form of Eq.(\ref{FSS}) then suggests that 
the linear dependence of $\langle N \rangle_P$ on the population size
gives $B =1$.
By arguments discussed below,
the FSS form~(\ref{FSS}) expresses the unified scaling law
for school sizes.
The scaling behavior of the school-size distribution is described by two
independent parameters:
the population density $\rho$ and
the breakup rate $p$.
One of the important results of the model is that a single parameter
$\langle N\rangle_P$ is sufficient to describe
the scaling behavior,
i.e.,
the two scaling relations of Eqs.(15) and~(16) are unified into
Eq.(\ref{FSS}) because of Eq.(13).

\begin{figure}[htbp]
 \centering
 \includegraphics[width=8cm]{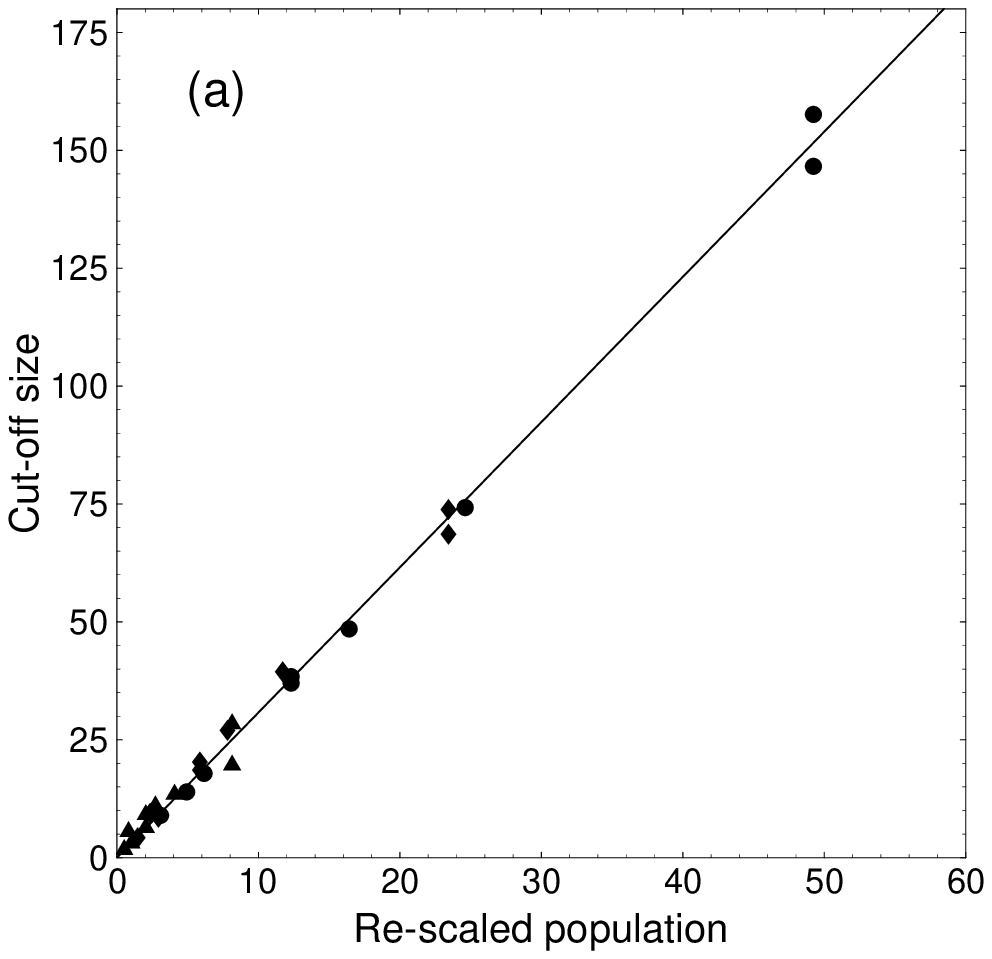}
 \includegraphics[width=8cm]{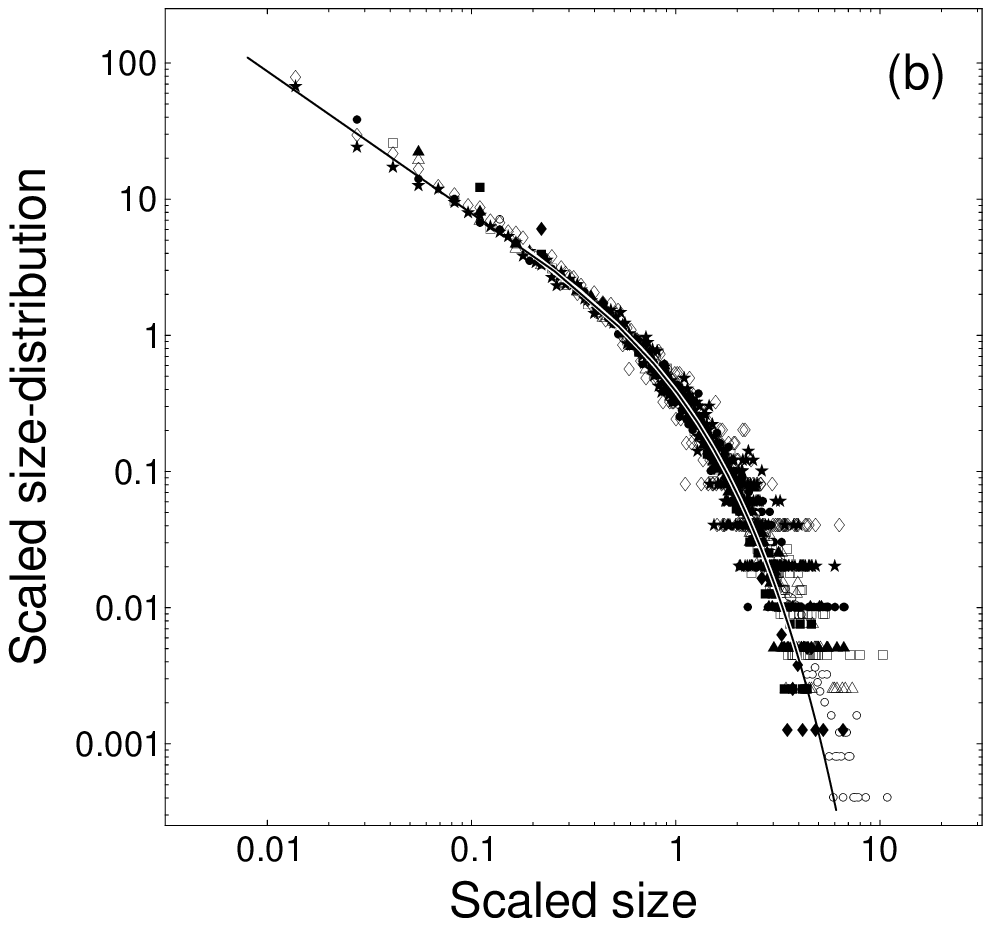}
 \caption{Unified scaling law for school sizes.
 (a)~$\langle N \rangle_P$ versus the re-scaled population size.
 The data in Fig.\ref{fig:3} are re-plotted in terms of re-scaled
 coordinate:
 the abscissa represents
 $(\lambda/\phi)\rho/p$.
 Factors $\lambda/\phi$ for
 one- ($\blacktriangle$),
 two-dimensional ($\blacklozenge$), and
 mean-field ($\bullet$) cases are estimated at
 0.163, 0.469, and 0.985, 
 respectively.
 The solid line is fit with slope 3.08.
 (b)~Simulated school-size distribution plotted on double-logarithmic
 scale (2D).
 The data in Figs.\ref{fig:4} and~\ref{fig:5} are re-plotted with
 $y= c\rho p^{-1} W(N)$
 as a function of the scaling variable
 $x=N/c\rho p^{-1}$,
 where $c=1.46$.
 The frequency distributions are normalized as
 $\sum_{N=1}^{\Phi} N W(N) = c\rho/p$.
 The re-scaled distributions collapse onto one another.
 The data collapse is equivalent to the FSS plot with
 $A =B =1$ in Eq.(\ref{FSS}).
 The solid line is a prediction of the mean-field theory \citep{Niwa03}.
 }
 \label{fig:6}
\end{figure}

To understand the unified scaling law for school sizes,
it is essential to see what determines the cut-off size of power law.
The cut-off size is determined by the competition between breakup and
coalescence of schools.
The coalescence is regulated by the spatial number-density of schools
resulting from a balance of the population density and the breakup rate
of schools.
Equation~(13) is then fundamental to unify the two scaling laws~(15)
and~(16).
Let us trace the size change of the school a certain individual (named
``A'') rides.
Let $\phi$ be the coarse-grained density of schools (i.e. number of
schools per site), and,
$\lambda$ the probability for a school to coalesce with other schools
per time step (i.e. coalescence rate).
Then the expected increment and decrement of the size of ``A''-riding
school (denoted by $N_{\mbox{\scriptsize A}}$) are
$c_0 (\rho/\phi) \lambda$
(accompanied by a constant factor $c_0$)
and
$p N_{\mbox{\scriptsize A}}/2$,
respectively, at each time step of coarse-grained simulation
[$\rho/\phi$ reads an (apparent) average size of fish schools].
The expected size-change of ``A''-riding school per time step,
$\Delta N_{\mbox{\scriptsize A}}$,
obeys
\begin{equation}
 \Delta N_{\mbox{\scriptsize A}}
  =
  c_0 \left(\frac{\rho}{\phi}\right)
  \lambda
  -
  \frac{p}2 N_{\mbox{\scriptsize A}}.
\end{equation}
The ratio
 $(\lambda c_0 \rho/\phi)/(p/2) $,
therefore, gives the expected size of ``A''-riding school in the
 stationary state,
which is equivalent to the value of $\langle N \rangle_P$ by definition.
Thus,
the fundamental equation~(13) is approved:
\begin{equation}
 \langle N \rangle_P
  =
  2 c_0
  \frac{\lambda}{\phi}
  \cdot
  \frac{\rho}{p},
\end{equation}
where $c_0$ is just a number independent of space dimension,
fit giving $1.54\pm0.02$ in the simulations (Fig.\ref{fig:6}a).
The values of the ratio of $c$ to $\lambda/\phi$ for
one-, two-dimensional, and mean-field cases are estimated at
3.09, 3.11, and 3.07, respectively,
which all exhibit the similar value independent of the case of the
model.
The space dimension is irrelevant to the (reduced) prefactor $c_0$.
(How does one a priori obtain the specific value of the dimensionless
quantity $c_0$ within the model setting?)
%

Thus the above two scaling relations can be unified and reduced to
\begin{equation}
 W(N;\rho,p) \mbox{d}N
  =
  N^{-1}
  G\left(
    N/ (2c_0 \lambda\phi^{-1}) \rho p^{-1}
  \right)
  \mbox{d}
  \left(
   N/(2c_0 \lambda\phi^{-1}) \rho p^{-1}
 \right),
\end{equation}
with normalization
\(
 \sum_{N=1}^\Phi N W (N)
 =
 (2c_0 \lambda\phi^{-1}) \rho p^{-1}
\),
which corresponding to Eq.(17).
A scaling function $G(x)$ has a strong drop for $x>1$.
To verify this law,
the simulated data in Figs.\ref{fig:4} and~\ref{fig:5} are re-plotted in
terms of re-scaled coordinates where the $x$-axis is chosen as
$x = N/ (2c_0 \lambda\phi^{-1}) \rho p^{-1}$,
and the $y$-axis represents
$y =
 (2c_0 \lambda\phi^{-1}) \rho p^{-1} W(N)$.
The resulting graph is shown in Fig.\ref{fig:6}b.
The data collapse is very good, implying a unified law for school sizes.

The same numerical results can be obtained with one-dimensional case
of the model.
Besides, the data-collapse in its mean-field case is depicted in
Fig.4 in \citet{Niwa03}.
Therefore, the unified scaling law, Eq.(20), holds independently of the
space dimension:
space does not influence a linear power-law behavior with a crossover to
an exponential decay around $\langle N \rangle_P$.
Figure~\ref{fig:7} in the next section showing the comparison of the
empirical school-size distribution with simulated distributions in one-,
two-dimensional, and mean-field cases directly verifies the irrelevance
of space in the scaling law for fish school sizes.

\section{The Integrity of Schools}

The cut-off size in the scaling law results from the ability of a school
to maintain its integrity over only a certain amount of time:
$\langle N \rangle_P$ varies in proportion as the coalescence rate, and
inversely as the breakup rate [see Eq.(19)].
The proposed numerical approach to fish school-size distribution may
answer yet another challenging question: how schools split.

Let us analyze the school sizes of Japanese sardine acoustically
surveyed in the summer in 1982 (the same data set as Fig.\ref{fig:1}).
Transect sampling was carried out for a length of $\xi =$ 1211.208
kilometers in a survey area of $\omega = 26904\;\mbox{km}^2$,
and 1522 schools (denoted by $\mu$) were observed.
The average (horizontal) interception-length of schools was
 $\overline{\ell} =$ 20.9 meters.
We now apply Buffon's needle problem to the estimation of the
dimensional size and the spatial number-density of fish schools
\citep{Doi79}.

Assume the shape of fish schools be horizontally disk-like with the
(average) diameter $\overline{R}$.
Let $a$ be the average of inter-school distance,
so that the number density of schools, $\tilde\phi$, is given by $a^{-2}$.
Imagine a disk (instead of a needle) dropping on a lined sheet of paper.
The probability of the disk hitting one of the lines is
$P_{\mbox{\scriptsize disk}} =\overline{R}/a$.
Then the average interception-length of schools is calculated as
$\overline{\ell} = {\pi (\overline{R}/2)^2}/a$.
Since we expect to observe one school for a transect sample interval of
length $\delta = \xi/\mu$, the average length of successive
detections of schools, 
the strip of width $a$ and length $\delta$ contains
$1/P_{\mbox{\scriptsize disk}}$ schools, i.e.
$\tilde\phi a\delta = 1/P_{\mbox{\scriptsize disk}}$.
Accordingly,
$\delta/a = a/\overline{R}$ is obtained.
Then the average diameter of schools and the average inter-school
distance are given by
\begin{equation}
 \overline{R}
  =
  \sqrt[3]{\frac{16}{\pi^2} \overline{\ell}^2 \delta}
  =
  82.6\; \mbox{m},
\end{equation}
and
\begin{equation}
 a
  =
  \sqrt[3]{\frac{4}{\pi} \overline{\ell} \delta^2}
  =
  256.4\; \mbox{m},
\end{equation}
respectively.
Thus the fine-grained spatial number-density of schools, $\tilde\phi$,
is estimated at
$15.2\; \mbox{km}^{-2}$.

We now return to the coarse-grained model of school aggregation.
Let the average diameter of schools, $\overline{R}$, be the lattice
constant.
The data are normalized to the coarse-grained density of schools,
\(
 \phi
  =
  \tilde\phi \overline{R}^2
  =
  0.104
\):
\begin{equation}
 s^{-1} \sum_{i} {\hat W}_i
  =
  \phi,
\end{equation}
where $s=2^{22}$ sites ($\approx \omega/\overline{R}^2$).
Then the coarse-grained population density is estimated at
\begin{equation}
 \rho
  =
  s^{-1} \sum_{i} N_i {\hat W}_i
  =
  0.415,
\end{equation}
where unit is half the class width of size histogram (per site).
The empirical data yield
$ \langle N \rangle_P = 11.7$.

%
\begin{table}[tb]
\caption[simulation]{
 {\bf Summary of simulations for Japanese sardine schools.}
Monte-Carlo simulations have been performed with coarse-grained zones of
 $s =2^{22}$ sites, $\Phi = 1.74\times 10^6$ individuals,
 and simulation run $=2^{17}$ time steps.
}
\vspace*{0.3cm}
\label{tab4}
\begin{center}
\renewcommand{\arraystretch}{1.2}
{
\begin{tabular*}{33pc}{lcccccc}
\noalign{\hrule height0.8pt}
 & $p$ & $\phi$ $^{\mbox{\scriptsize a}}$ 
 & $\langle N \rangle_P$\/\ $^{\mbox{\scriptsize a}}$ & half-life $T$ $^{\mbox{\scriptsize b}}$
 & $\lambda$ & $p_T\; (=p\cdot T)$\\
\noalign{\hrule height0.8pt}
1D ($\vartriangle$) & 0.0179 & 0.105 & 11.7 & 23.5 & $1.15\times 10^{-2}$ & 0.421\\
2D ($\lozenge$) & 0.0516 & 0.092 & 13.6 & 7.55 & $4.02\times 10^{-2}$ & 0.390\\
mf ($\circ$) & 0.107 & 0.109 & 10.9 & 3.45 (3.29)$^{\mbox{\scriptsize c}}$
 & $9.41\times 10^{-2}$ & 0.369\\
\noalign{\hrule height0.8pt}
\multicolumn{7}{l}{$^{\mbox{\scriptsize a}}$computed from simulation
 results at the last of run}\\
\multicolumn{7}{l}{$^{\mbox{\scriptsize b}}$Unit is simulation time step.}\\
\multicolumn{7}{l}{$^{\mbox{\scriptsize c}}$predicted by
 $(\ln 2)/(p+\phi)$ with $\phi=0.104$
 (evaluated value from empirical data)}\\
\end{tabular*}
}
\end{center}
\end{table}
%
\begin{figure}[tb]
 \centering
 \includegraphics[width=8cm]{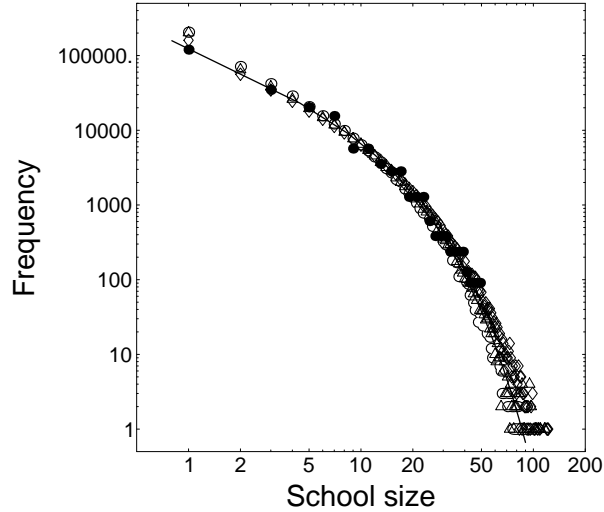}
 \caption{
Empirical school-size distribution of sardine observed in the summer in
 1982 ($\bullet$),
 same as Fig.\ref{fig:1} but with normalization Eq.(23),
 compared with
 simulated distributions after $2^{17}$ time steps in
 one- ($\vartriangle$),
 two-dimensional ($\lozenge$), and
 mean-field ($\circ$) cases of the model.
 The simulated distributions are in good agreement with empirical data.
 Different plots for the empirical observation and the numerical
 simulations fall onto  one another.
 The solid line is a prediction of the mean-field theory 
 [Eq.(11) in \citet{Niwa03}],
 with normalization
 \(
 \Phi\cdot
 \left[ \langle N \rangle_P
 \int_0^{\infty} \exp
 \left[-x \left( 1-e^{-x}/2 \right) \right]\mbox{d}x
 \right]^{-1}
 =
 1.31\times 10^5
 \),
 consistent with Eq.(24),
 where $\langle N \rangle_P = 11.7$.
 The simulation parameters are summarized in Table~\ref{tab4}.
 Plotted on double-logarithmic scale.
 }
 \label{fig:7}
\end{figure}
Now we can numerically experiment on the school size distribution with
the population $\Phi = s\rho$
and the breakup rate $p$ listed in Table~\ref{tab4}.
The breakup probability $p$ at each simulation time step is estimated at
$c\rho/\langle N \rangle_P$, by using Eq.(13) with prefactors
resulting from simulations (Fig.\ref{fig:3}).
A one-sized school of simulated fish ($N=1$) is considered as an atomic
object, which contain a certain number of fish correspond to half the
class width of frequency data of sardine school sizes observed in the
wild.
Starting from random initial configurations of eight-sized schools on
the lattice,
the coarse-grained simulations imitate the fine-grained situation of the
real-world, as shown in Fig.\ref{fig:7}.
The remarkable consistency between the empirical data and the model's
prediction unambiguously describes that
the space dimension cannot be relevant to the scaling exponent of
power-law school-size distribution.
%

%
\begin{figure}[tb]
 \centering
 \includegraphics[width=8cm]{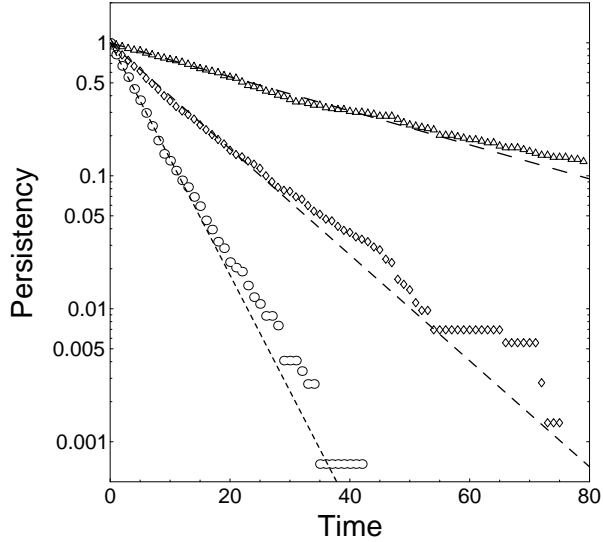}
 \caption{
 Persistency of simulated sardine schools in
 one- ($\vartriangle$),
 two-dimensional ($\lozenge$), and
 mean-field ($\circ$) cases of the model.
 Shown is a ratio of the number of schools maintaining their integrity
 over a certain amount of time steps of simulation run without
 coalescing or splitting.
 Broken lines depict the exponential decays with rates
 $p+\lambda = 0.0294$ (1D),
 $0.0918$ (2D), and $0.201$ (mf)
 fitting the numerical results, respectively
 (unit: per time step).
 The simulation parameters are summarized in Table~\ref{tab4}.
 Plotted on semilogarithmic scale.
}
 \label{fig:8}
\end{figure}
Let us investigate the dynamic properties from the numerical results.
There are two competing processes in the interacting school system:
breakup and coalescence.
The disintegration behavior of fish schools consists of these two
processes.
The breakup and coalescence rates, therefore, combine into the
probability of disintegration per unit time per school, which gives the
half-life $T$ as its reciprocal:
\begin{equation}
 T =
  \frac{\ln 2}{p+\lambda}.
\end{equation}
After time $T$,
the number of schools maintaining their integrity is half of the
original number (Fig.\ref{fig:8}).
In the mean-field case of the model, the coalescence probability per
simulation time step is given by the density of schools,
$\lambda = \phi$,
which is tested numerically.
The half-life of simulated fish schools is calculated in Table~\ref{tab4}.
The breakup probability per half-life, $p_T\;( = p\cdot T)$,
obtained for each case (1D/2D/mf)
exhibits the similar value independent of the case of the
model.
Therefore, if the half-life of sardine school is experimentally
determined in the wild
[e.g. \citet{Lester-etal85} or \citet{Bayliff88} for skipjack tuna],
we can estimate the breakup rate and the coalescence rate in the wild at
$p_T/T$ and $(\ln 2 - p_T)/T$, respectively,
with numerically predicted value $p_T$.

\section{Discussions}

Data collapse is a way of establishing the scaling in fish school-size
distributions.
I have extracted the power-law exponent of size distributions via a
minimization of a measure to quantify the nature of finite-size scaling
collapse,
in contrast to the `best-by-eye' data-collapse method.
The number $W(N)$ of $N$-sized schools decays with the truncated
power-law form with exponent $-1$, and the power-law cut-off scales with
$\langle N \rangle_P$,
which is a well-defined mean of the population distribution among school
sizes, $NW(N)$.
In order to explain the observed scaling property,
I have chosen a stationary equilibrium model,
contrary to a stationary non-equilibrium system of Bonabeau et al.'s
(1995, 1998, 1999) model.
It has been found that
the scaling law for school sizes of pelagic fish is completely
independent of the dimension of the space in which the fish move:
the space dimension is irrelevant to the power-law exponent of size
distributions.
This result is contrary to common knowledge in physics that the critical
exponents characterizing the scaling behavior observed in phase
transitions depend on the dimensionality.
The model of school aggregation does not perfectly mirror the
interacting school system in the wild
(e.g. the size of a school may be relevant to its breakup probability),
yet the numerically simulated result conforms almost perfectly to the
empirical data.
This is a consequence of universality \citep{Stanley95}.

The scaling law supports the view that
the power-law distribution of fish schools is a self-organized critical
phenomenon (Bak, 1996; Jensen, 1998),
not merely a reflection of an exponential distribution of population among
school sizes,
because only critical processes exhibit data-collapse (Yeomans, 1992;
Bak {et al.}, 2002, Christensen {et al.}, 2002), known as
scaling in critical phenomena.
The interacting school system is naturally attracted to the critical
value of the spatial number-density of fish schools,
without any external adjustment being necessary.
In the system there are two competing processes, coalescence and
breakup, and the critical density depends crucially on the interplay
between the coalescence and breakup time scales.
If the school density becomes greater (or less) than the critical value,
this increased (or decreased) density in turn increases (or decreases)
the probability of coalescence, leading to a shift toward less (or
greater) density until it reverts to the critical value.

The relation between dimensional and social sizes of pelagic fish
schools, Eq.(1), gives another scaling law for the school size,
which is analogous to that used in polymer physics \citep{deGennes79}.
What is universal in this law is the exponent $\nu$:
it is the same for all schools, supposed to be independent of not only
environmental conditions but also species.
What is non-universal here is the prefactor.
It depends on the details of interactions between individuals.
If we want to understand the properties of schooling configurations, the
first step is to explain the existence and the value of the exponent
$\nu$.
The second step is to account for the constant that multiplies
$N^{\nu}$,
and this involves delicate studies on local properties within a school.
In the first stage, the school size $N$ (in number or biomass) must be
measured for different values of dimensional size $R$ and compare them.
Finding the exact value of $N$ for a given school is not an easy task,
whereas acoustic surveys for pelagic species are extensively performed. 

By contraries, the exponent $\nu$ may be estimated
by hypothesizing FSS in the school-size distribution of pelagic species.
Assume the scaling relation, Eq.(1), and 
choose the suitable value of $\nu$ to achieve the best data-collapse in
the FSS analysis,
while a dimensional size histogram for acoustic-survey data is
transformed into a social size histogram by using Eq.(2).
Once the scaling relation between dimensional and social sizes of
pelagic fish schools is established, the precision in the stock
assessment will be largely improved.


%

\end{document}